\documentclass[12pt,english]{article}

\usepackage[T1]{fontenc}
\usepackage[latin9]{inputenc}
\usepackage[dvipsnames]{xcolor}
\usepackage{babel}
\usepackage{amsmath}
\usepackage{amsthm}
\usepackage{amssymb}
\usepackage{mathtools}
\usepackage{graphicx}
\usepackage{subcaption}
\usepackage{color}
\usepackage{soul}
\usepackage[mathscr]{euscript}

\usepackage{titlesec}

\titleformat{\paragraph}
{\normalfont\normalsize\bfseries}
{\theparagraph}{1em}{}

\titlespacing*{\paragraph}
{0pt}{1ex}{1ex}

\makeatletter

\usepackage{jheppub}
\def\@fpheader{\relax}
\allowdisplaybreaks[1]

\makeatother

\begin{document}

\title{Effective Improved-GUP Cosmology: Emergent FLRW Universe without a Bounce}


\author[a,b,c]{Saeed Rastgoo,}
\author[d,e]{Wilfredo Yupanqui}
\affiliation[a]{Department of Physics, University of Alberta, Edmonton, Alberta T6G 2G1, Canada} 
\affiliation[b]{Department of Mathematical and Statistical Sciences, University of Alberta, Edmonton, Alberta T6G 2G1, Canada}
\affiliation[c]{Theoretical Physics Institute, University of Alberta, Edmonton, Alberta T6G 2G1, Canada}
\affiliation[d]{Unidad Acad\'emica de F\'isica, Universidad Aut\'onoma de Zacatecas, Calzada Solidaridad esquina con Paseo a la Bufa S/N, Zacatecas, 98060, Zacatecas, M\'exico}
\affiliation[e]{Departamento de F\'isica, Divisi\'on de Ciencias e Ingenier\'ias, Universidad de Guanajuato, Loma del Bosque 103, Le\'on, 37150, Guanajuato, M\'exico}
\emailAdd{srastgoo@ualberta.ca}
\emailAdd{w.yupanqui@fisica.uaz.edu.mx}









\abstract{
We investigate the effective dynamics of a spatially flat FLRW universe coupled to a massless scalar field by applying the improved generalized uncertainty principle (GUP)-inspired deformations to the algebra of Ashtekar-Barbero variables. We consider deformations, once only in the geometry sector, and once in both the geometry and matter sectors. We show that in both cases, the classical singularity is replaced with a non-singular, emergent universe coasting from a constant-volume state in the infinite relational past time without a bounce. As expected, a constant geometric GUP parameter leads to fiducial anomalies, but we resolve it by employing an improved scheme in which this parameter is inversely proportional to the gravitational momentum. This leads to a universal, invariant maximum energy density while preserving the emergent nature of the universe. A Lyapunov stability analysis reveals that the improved scheme drives the universe toward classicality faster than the constant scheme.
}







\maketitle

\section{Introduction}

The singularity theorems of Penrose and Hawking establish that classical
general relativity inevitably predicts the breakdown of spacetime
at extreme curvatures, most notably in the Big Bang. This singularity
is widely understood to signal the failure of the classical theory
rather than a physical feature of nature, and its resolution is expected
to emerge from a consistent quantum theory of gravity. In the absence
of such a complete theory, approaches that incorporate key quantum
gravitational effects into the classical framework have proven to
be a productive avenue of investigation. Two of the most developed
such approaches are loop quantum cosmology (LQC) and models based
on the generalized uncertainty principle (GUP).

LQC modifies the dynamics of homogeneous cosmologies by replacing
the classical Hamiltonian with a discrete, polymer-quantized analogue,
typically leading to a quantum bounce that connects a contracting
pre-Big Bang branch to the current expanding universe \cite{Bojowald:2001xe, Ashtekar:2003hd, Ashtekar:2006rx}.
GUP, on the other hand, encodes a minimum length scale directly into
the phase space algebra, deforming it by a small parameter typically
of the order of the Planck length \cite{G.Veneziano_1986,MAGGIORE199365,Kempf:1994su,SCARDIGLI199939,PhysRevD.105.L121501,BIZET2023137636}. GUP has been
applied to a variety of gravitational settings, including black holes,
thermodynamics, and cosmology \cite{Adler:2001vs,Das:2008kaa, Hossenfelder:2012jw,YupanquiCarpio2025}. 

A critical nuance in effective quantum cosmology is the treatment
of the fiducial cell. In homogeneous models, a fiducial cell of volume
$\mathring{v}$ must be introduced to render symplectic integrations
finite. Physical observables must remain independent of $\mathring{v}$
and be invariant under the arbitrary rescaling of it. An important
development in both LQC \cite{Ashtekar2006_mubar} and GUP \cite{Fragomeno:2024tlh, Bosso:2023fnb, Blanchette:2021vid, Bosso:2020ztk, Blanchette:2020kkk, Rastgoo:2022mks} has been the recognition that the
choice of how the deformation parameter depends on the dynamical variables
--- the so-called improved schemes --- has profound physical consequences,
in particular for the resolution of fiducial-cell artifacts and singularities. 

The application of GUP ideas to cosmology has rather a long history,
approached from several distinct angles. In one of the earliest direct
applications to homogeneous cosmological minisuperspaces \cite{Battisti:2007vt, PhysRevD.77.023518},
using a massless scalar field as a relational clock --- the same
strategy we adopt here --- the authors found that the classical singularity
is probabilistically suppressed and no Big Bounce appears. This no-bounce
result was confirmed in the Bianchi I and Taub models \cite{Vakili:2007zz}.
A parallel and complementary line of research derived GUP-modified
Friedmann equations not from a phase space deformation but from the
thermodynamics of cosmological horizons: by modifying the entropy-area
relation of the FLRW apparent horizon to incorporate the minimal length,
one recovers modified Friedmann equations featuring a maximum energy
density and the potential absence of the initial singularity \cite{Zhu:2008cj, Cai:2008ys}.
This thermodynamic route has been extended to linear, quadratic, and
higher-order GUP formulations, including extended uncertainty principles
with both UV and IR cutoffs \cite{Giardino:2021}. GUP corrections
to inflationary cosmology have also been investigated extensively,
where a minimum length sets a natural trans-Planckian cutoff on the
primordial spectrum \cite{Hassan:2002, Ashoorioon:2005}. More recently,
the deformation of the cosmological phase space Poisson algebra has
been applied to quintessence and phantom scalar field cosmologies,
revealing new fixed points and altered inflationary dynamics \cite{Paliathanasis_2015, Bhandari2024}.
The sign of the GUP deformation parameter has been recognized as physically
decisive in several of these works, with negative values more naturally
compatible with singularity resolution \cite{Battisti:2007vt,ATAZADEH201787, Fragomeno:2024tlh, Gingrich:2024mgk}.
For a comprehensive review of the field, including constraints from
atomic, gravitational wave, and cosmological observations, see \cite{Bosso:2023, Hossenfelder:2012jw}. 

In this paper, we construct an effective cosmological model by introducing
GUP-inspired deformations into the canonical algebra of the Ashtekar-Barbero
variables $(c,p)$ and the matter sector $(\phi,p_{\phi})$. Our objective
is twofold: first, to determine the exact dynamical trajectory of
the universe under these deformations, and second, to construct a
physically viable improved scheme that strictly respects fiducial
invariance. We treat the problem systematically by considering two
main deformation cases: one in which only the geometric sector algebra
is modified, and one in which both the geometric and matter sector
algebras are deformed. Within the first case, we study a constant
deformation parameter and an improved momentum-dependent one, drawing
a close parallel with the $\mu_{0}$ and $\bar{\mu}$ schemes in LQC.
Unlike LQC, however, we find that GUP corrections do not produce a
quantum bounce. Instead, for a negative deformation parameter, the
classical Big Bang singularity is replaced by a non-singular emergent
phase in which the universe asymptotically coasts out of a constant-volume
state in the infinite past. This picture is reminiscent of the emergent
universe scenario of Ellis and Maartens \cite{Ellis:2003qz, Ellis:2003ue},
and we make this connection explicit through a fixed-point and stability
analysis. This analysis shows that instabilities under perturbations
``repel'' the universe out of the constant-volume phase into expansion.
Thus, the present work differs from previous body of literature in
several respects: we work in the Ashtekar-Barbero connection variables
rather than the metric minisuperspace variables and we implement an
improved momentum-dependent scheme (which is never studied before)
to eliminate the fiducial-cell anomaly. Moreover, we derive exact
closed-form solutions using the scalar field as a relational clock
in a systematic manner, and also provide a systematic stability analysis
of the emergent phase through Lyapunov exponents.

The structure of the paper is as follows. In Section \ref{Sec:Class-dynam}
we review the classical dynamics of the system and relevant quantities.
In Section \ref{Sec:deform-geometry} we introduce the GUP deformations
in the geometry sector, leading the matter sector unmodified. We study
two subcases in one of which the GUP parameter $\beta$ is constant
and in the other one, it is inversely proportional to momentum, which
is the so-called improved case. In Sec. \ref{subsec:Deform-both},
we simultaneously deform both the geometry and matter sectors and
study the consequences. Finally, in Sec. \ref{Sec:Conclusion} we
summarize and conclude.

\section{Classical dynamics\label{Sec:Class-dynam}}

We consider a spatially flat FLRW universe with a metric
\begin{equation}
ds^{2}=-N^{2}dt^{2}+a^{2}(t)\left(dx^{2}_{1}+dx^{2}_{2}+dx^{2}_{3}\right)
\end{equation}
without a cosmological constant coupled to a massless scalar field.
Throughout the text we will make use of both coordinate time derivative
$\dot{F}=\tfrac{dF}{dt}$ and proper time derivative $F^{\prime}=\tfrac{dF}{d\tau}$
where the proper time $\tau$ is defined from the above metric as
$d\tau=Ndt$. This means that $F^{\prime}=N^{-1}\dot{F}$.

The geometry sector of this system can be canonically described, using
the symmetry-reduced Ashtekar-Barbero variables, by the phase space
variables \cite{PhysRevD.73.124038}
\begin{align}
c= & \gamma a^{\prime}\mathring{v}^{\frac{1}{3}}=\frac{\gamma\dot{a}}{N}\mathring{v}^{\frac{1}{3}}, & \left|p\right|= & a^{2}\mathring{v}^{\frac{2}{3}}\label{eq:c-p-def}
\end{align}
where $p$ is the momentum conjugate to $c$, $a$ is the scale factor,
and $\gamma$ is the constant Barbero-Immirzi parameter. Here $c$
is the only component of the Ashtekar connection and $p$ is the only
component of the densitized triad. Moreover, $\mathring{v}$ is the
volume of a fiducial spatial cell which is introduced to make the
symplectic structure well-defined in homogeneous models \cite{PhysRevD.73.124038}.
This volume can be considered a scaling factor and can be chosen arbitrarily.
The matter is described by the scalar field $\phi$ and its conjugate
momentum $p_{\phi}$. 

The classical dynamics of this universe can be described by the Hamiltonian
\begin{equation}
H=N\mathcal{C}=N\left(-\frac{3}{\kappa\gamma^{2}}c^{2}\left|p\right|^{\frac{1}{2}}+\frac{1}{2}\frac{p^{2}_{\phi}}{\left|p\right|^{\frac{3}{2}}}\right),\label{eq:Hamilton_const_C}
\end{equation}
where $\mathcal{C}$ is the Hamiltonian constraint \cite{PhysRevD.78.064072}
and $\kappa=8\pi G$. The canonical algebra of the four-dimensional
phase space constructed out of $(c,p;\phi,p_{\phi})$ reads 
\begin{align}
\left\{ c,p\right\} = & \frac{\kappa\gamma}{3}, & \left\{ \phi,p_{\phi}\right\} = & 1.\label{eq:PB-orig-cp}
\end{align}
Taking advantage of the fact that $\mathcal{C}$ is a first class
constraint and $N$ is its associated Lagrange multiplier whose choice
does not affect the classical physics, we set 
\begin{equation}
N=\mathcal{A}\left|p\right|^{\frac{3}{2}}=\mathcal{A}a^{3}\mathring{v},\label{eq:Classical-Lapse}
\end{equation}
with $\mathcal{A}$ being a dimensionful constant, to simplify $H$
in (\ref{eq:Hamilton_const_C}) as 
\begin{equation}
H=N\mathcal{C}=-\frac{3\mathcal{A}}{\gamma^{2}\kappa}c^{2}p^{2}+\frac{\mathcal{A}}{2}p^{2}_{\phi}.\label{eq:Class_Hamilt_reduc}
\end{equation}
In this work, we adopt natural units by setting the speed of light
$v_{c}=1$ and $\hbar=1$. Within this framework, $\left[G\right]=\text{L}^{2}$.
Moreover, from the $t$ part of the metric we see that $\left[N^{2}dt^{2}\right]=\text{L}^{2}$.
Since coordinate time carries dimensions of length, $\left[dt\right]=\text{L}$,
we find $\left[N\right]=1$. Using these and (\ref{eq:c-p-def}) and
since $\left[a\right]=1=\left[\gamma\right]$, we obtain $\left[p\right]=\text{L}^{2}$
and $\left[c\right]=1$. Based on these and the choice of lapse (\ref{eq:Classical-Lapse})
we find $\left[\mathcal{A}\right]=\text{L}^{-3}$. From this and the
fact that $\left[H\right]=\text{L}^{-1}$ we get $\text{\ensuremath{\left[p_{\phi}\right]}}=\text{L}$
and due to the algebra of $\phi$ and $p_{\phi}$ one deduces $\left[\phi\right]=\text{L}^{-1}$.

Using the Hamiltonian (\ref{eq:Class_Hamilt_reduc}) and the canonical
algebra (\ref{eq:PB-orig-cp}), one can derive the classical equations
of motion (EoM) as
\begin{align}
c^{\prime}=\frac{1}{N}\dot{c}=\frac{1}{N}\left\{ c,H\right\} = & -\frac{2}{\gamma}\frac{\text{sgn}\left(p\right)}{\left|p\right|^{\frac{1}{2}}}c^{2},\label{eq:Class-EoM-c-tau}\\
p^{\prime}=\frac{1}{N}\dot{p}=\frac{1}{N}\left\{ p,H\right\} = & \frac{2}{\gamma}\left|p\right|^{\frac{1}{2}}c,\label{eq:Class-EoM-p-tau}\\
\phi^{\prime}=\frac{1}{N}\dot{\phi}=\frac{1}{N}\left\{ \phi,H\right\} = & \frac{p_{\phi}}{\left|p\right|^{\frac{3}{2}}},\label{eq:Class-EoM-phi-tau}\\
p^{\prime}_{\phi}=\frac{1}{N}\dot{p}_{\phi}=\frac{1}{N}\left\{ p_{\phi},H\right\} = & 0.\label{eq:Class-EoM-pphi-tau}
\end{align}
It is immediately seen from these EoM that $p_{\phi}$ is a constant
of motion and a Dirac observable, which is expected as $\phi$ does
not appear in the Hamiltonian (\ref{eq:Class_Hamilt_reduc}).

The equations of motion (\ref{eq:Class-EoM-c-tau}) and (\ref{eq:Class-EoM-p-tau})
are equivalent to the Friedmann equations for this cosmology when
written in terms of $a$ and $a^{\prime}$. To see this, one can notice
from (\ref{eq:c-p-def}) that
\begin{equation}
\frac{p^{\prime}}{2p}=\frac{a^{\prime}}{a}=H\label{eq:H-in-p}
\end{equation}
Using (\ref{eq:Class-EoM-p-tau}) in the above equation, and then
replacing for $c$ from the Hamiltonian constraint (\ref{eq:Hamilton_const_C})
using $\mathcal{C}\approx0$ as
\begin{equation}
c=\pm\sqrt{\frac{\kappa}{6}}\gamma\frac{p_{\phi}}{\left|p\right|},\label{eq:c-from-constraint}
\end{equation}
in the result yields
\begin{equation}
\left(\frac{a^{\prime}}{a}\right)^{2}=H^{2}=\left(\frac{p^{\prime}}{2p}\right)^{2}=\frac{1}{2}\frac{p^{2}_{\phi}}{\left|p\right|^{3}}=\frac{\kappa}{3}\left(\frac{1}{2}\frac{p^{2}_{\phi}}{a^{6}\mathring{v}^{2}}\right)=\frac{\kappa}{3}\rho\label{eq:1st-Fied-class}
\end{equation}
Moreover we can obtain the second Friedmann equation by using the
equations of motion, definitions (\ref{eq:c-p-def}) and (\ref{eq:c-from-constraint})
to obtain
\begin{equation}
\frac{a^{\prime\prime}}{a}=H^{\prime}+H^{2}=-2H^{2}=-\frac{2}{\gamma^{2}}\frac{c^{2}}{\left|p\right|}=-\frac{\kappa}{6}\left(\frac{2p^{2}_{\phi}}{\left|p\right|^{3}}\right)=-\frac{\kappa}{6}\left(\frac{2p^{2}_{\phi}}{a^{6}\mathring{v}^{2}}\right)=-\frac{\kappa}{6}\left(\rho+3P\right)\label{eq:Second_Fried_eq_usuall}
\end{equation}
These correspond to a spatially flat FLRW universe ($k=0$) without
a cosmological constant, coupled to a massless scalar field (stiff
matter) with energy density $\rho$ and pressure $P$,
\begin{align}
\rho= & \frac{1}{2}\frac{p^{2}_{\phi}}{\left|p\right|^{3}}=\frac{1}{2}\frac{p^{2}_{\phi}}{a^{6}\mathring{v}^{2}}, & P= & \rho,\label{eq:rho-p-class}
\end{align}
where we have used the equation of state $P=w\rho$ with $w=1$ for
the massless scalar field. 

Let us proceed to find the classical solutions to the EoM. In canonical
gravity, the coordinate time is a gauge parameter with respect to
which all observables are constants of motion. So one chooses a physical
clock variable in this case $\phi$ and find the evolution with respect
to that. To proceed with this, we first note that 
\begin{equation}
V=a^{3}\mathring{v},\,\left|p\right|=\mathring{v}^{\frac{2}{3}}a^{2}\Rightarrow\left|p\right|=V^{\frac{2}{3}}\label{eq:p-vs-V}
\end{equation}
Thus if we find the evolution of $p$ with respect to $\phi$ rather
than $t$ or $\tau$, we can study the evolution of the universe's
volume as a function of the physical clock $V(\phi)$. To this end,
we use (\ref{eq:Class-EoM-p-tau}) and (\ref{eq:Class-EoM-phi-tau})
to obtain
\begin{equation}
\frac{p^{\prime}}{\phi^{\prime}}=\frac{dp}{d\phi}=\frac{2}{\gamma}\frac{p^{2}c}{p_{\phi}}.\label{eq:dpdphi-class}
\end{equation}
Replacing $c$ from (\ref{eq:c-from-constraint}) the above yields
\begin{equation}
\frac{dp}{d\phi}=\pm\sqrt{\frac{2\kappa}{3}}\left|p\right|.
\end{equation}
with the solution 
\begin{equation}
p\left(\phi\right)=p_{0}e^{\pm\text{sgn}\left(p_{0}\right)\sqrt{\frac{2\kappa}{3}}\phi},\label{eq:p-phi-class}
\end{equation}
where $p_{0}$ is a constant of integration and it is clear that $\text{sgn}\left(p\right)=\text{sgn}\left(p_{0}\right)$.
The two branches of this solution correspond to expanding and contracting
universes. Equivalently, one can write 
\begin{equation}
V=V_{0}e^{\pm\text{sgn}\left(p_{0}\right)\sqrt{\frac{3\kappa}{2}}\phi}\label{eq:V-phi-Class}
\end{equation}
where $V_{0}=\left(\text{sgn}\left(p_{0}\right)p_{0}\right)^{\frac{3}{2}}$
is a constant of integration corresponding to the volume of the universe
at $\phi=0$. The $+$ sign in (\ref{eq:p-phi-class}) and (\ref{eq:V-phi-Class})
corresponds to an expanding universe while the $-$ sign is associated
to a contracting one. Equation (\ref{eq:V-phi-Class}) shows that
the scalar field can be considered as a clock variable with respect
to which the volume of the universe changes. Hence, since classically
at $\tau=0$, have $V=0$ and $\rho\to\infty$ (from (\ref{eq:rho-p-class})),
we obtain the infamous classical singularity at the beginning (or
end) of the universe.

\begin{figure}[htb!] 
  \centering
  \includegraphics[width=0.5\linewidth]{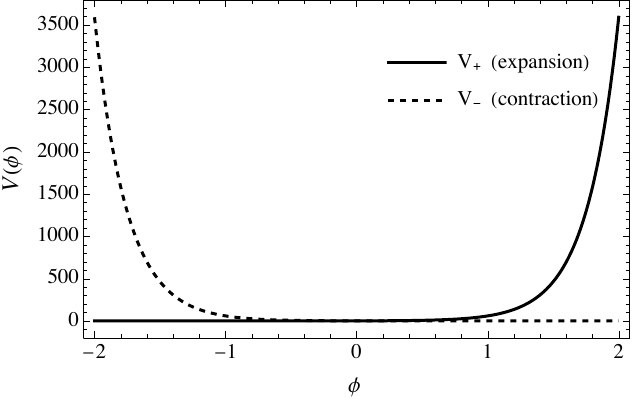} 
  \caption{Behavior of the classical \(V\) as a function of \(\phi\) in \eqref{eq:V-phi-Class}, 
with $\text{sgn}(p_{0})=+1$. The positive branch $V_{+}$ describes the 
expansion of the Universe, while the negative branch $V_{-}$ corresponds 
to its contraction.}
  \label{fig:Phi_vs_p_usuall} 
\end{figure}

Figure~\ref{fig:Phi_vs_p_usuall} displays the graphical representation of 
\eqref{eq:V-phi-Class}. The choice $\text{sgn}(p_{0})=+1$ ensures that 
$V_{0}$ remains real. The expanding phase, interpreted as a Big Bang, is 
depicted by the solid black curve, while the contracting phase (Big Crunch) 
is represented by the dashed curve.

\section{Effective Dynamics\label{Sec:Effective-dynamics}}

We will now introduce quantum gravity effects in this model by applying
an effective scheme based on improved generalized uncertainty principle
approach \cite{Fragomeno:2024tlh}. In this approach one deforms
the classical Poisson algebra (\ref{eq:PB-orig-cp}) by a momentum
dependent quantum parameter $\bar{\beta}\left(p\right)$. We will
study two cases. In the first case we only deform the algebra of the
geometric sector $(c,p)$. In the second case the algebra of both
the geometric and matter sector $(\phi,p_{\phi})$ are modified.

\subsection{Case 1: Deforming The Geometry Sector \label{Sec:deform-geometry}}

The deformation in the geometry section is done by modifying the classical
algebra (\ref{eq:PB-orig-cp}) as 
\begin{equation}
\left\{ c,p\right\} =\frac{\kappa\gamma}{3}f_{c}\left(\beta_{c},c\right),\label{eq:PB-orig-cp-GUP}
\end{equation}
where
\begin{equation}
f_{c}\left(\beta_{c},c\right)=\left(1+\beta_{c}c^{2}\right).\label{eq:fcp-def}
\end{equation}
Here, $\beta_{c}$ can be either a constant (the so called $\beta_{0}$
case in line with a similar naming in loop quantum gravity) or can
depend on the momentum as $\beta_{c}\left(p\right)$, i.e., the so
called improved scheme. In either case, $c$ and $p$ no longer form
a canonical pair. As a result, the dynamics is modified although the
Hamiltonian remains the same.

The particular choice of $f_c$ in \eqref{eq:fcp-def} is inspired by the 
Extended Uncertainty Principle (EUP) model, widely discussed in the 
literature~\cite{COSTAFILHO2016367,PARK2008698,10.1063/1.531501,Dabrowski2019,Hossenfelder:2012jw}. At the quantum level, this framework modifies 
the commutation relation between position and momentum as
\begin{equation}
    [\hat{c},\hat{p}]=i\frac{\kappa\gamma}{3}(1+\beta_c c^2),
\end{equation}
which leads to the modified uncertainty relation
\begin{equation}
    \Delta c\,\Delta p\geq\frac{\kappa\gamma}{6}\left[1+\beta_c(\Delta c)^2\right],
\end{equation}
and implies the existence of a minimal uncertainty in momentum of the order 
$(\Delta p)_{\text{min}}\propto \sqrt{\beta_c}$.

In contrast, within the GUP framework the typical choice is 
$f_c=1+\beta_c p^2$, which instead leads to a minimal uncertainty in the 
position variable~\cite{Kempf:1994su}. More generally, such constructions are 
referred to as generalized uncertainty relations.

In this work, we implement a deformation of the Poisson brackets between 
generalized coordinates and momenta in order to realize a minimal momentum. 
Therefore, our model is conceptually closer to the EUP framework, although 
we adopt well-established techniques developed in the context of GUP.

Notice that one can instead retain the
canonical algebra by making a nonlinear transformation of the connection
components as \cite{Bosso-2021,PhysRevD.97.126010}
\begin{equation}
    d\tilde{c}=\frac{dc}{1+\beta_{c}c^{2}},
\end{equation}
which leads to
\begin{equation}
    \tilde{c}=\frac{1}{\sqrt{\beta_{c}}}\tan^{-1}\left(\sqrt{\beta_{c}}c\right),
\end{equation}
for which
\begin{equation}
    \left\{ \tilde{c},p\right\} =\frac{\kappa\gamma}{3}.
\end{equation}
This way one works with the canonical variables $\tilde{c},p$ while
now the Hamiltonian is modified and written in terms of $\tilde{c}$.
We do not pursue this method and work with the modified algebra (\ref{eq:PB-orig-cp})
and the Hamiltonian constraint (\ref{eq:Hamilton_const_C}).

From the Hamiltonian (\ref{eq:Class_Hamilt_reduc}), and in this case
using the modified algebra (\ref{eq:PB-orig-cp-GUP}), we obtain the
effective equations of motion, given by 
\begin{align}
c^{\prime}= & -\frac{2}{\gamma}\frac{\text{sgn}\left(p\right)}{\left|p\right|^{\frac{1}{2}}}c^{2}f_{c},\label{eq:c-prime-eff-general}\\
p^{\prime}= & \frac{2}{\gamma}\left|p\right|^{\frac{1}{2}}cf_{c},\label{eq:p-prime-eff-general}\\
\phi^{\prime}= & \frac{p_{\phi}}{\left|p\right|^{\frac{3}{2}}},\label{eq:phi-prime-eff-general}\\
p^{\prime}_{\phi}= & 0.\label{eq:pphi-prime-eff-general}
\end{align}
As expected only the EoM for $c$ and $p$ are modified, while the
EoM for $\phi$ and $p_{\phi}$ remain unaffected. Using the same
method as before, from the EoM of $p$ and $\phi$ (and replacing
for $c$ from (\ref{eq:c-from-constraint})) we obtain
\begin{equation}
\frac{dp}{d\phi}=\pm\sqrt{\frac{2\kappa}{3}}\left|p\right|f_{c}.\label{eq:EoM-diff-eff}
\end{equation}
The solution to this differential equation depends on whether $f_{c}$
(or $\beta_{c}$) depends on $p$, and if so, with what functional
dependency. We will consider three subcases in the next sections.

Note that in the modified scheme we cannot use the definition of $c$
in (\ref{eq:c-p-def}). This is because that definition is dynamical
(depends on $a^{\prime}$) and hence it gets modified due to modified
dynamics of GUP. Instead we should use the modified definition of
$c$. This is obtained by taking the derivative of both sides of $p$
in (\ref{eq:c-p-def}), replacing the $p^{\prime}$ term using the
modified EoM (\ref{eq:p-prime-eff-general}), and using the definition
of $p$ in (\ref{eq:c-p-def}) again. The new definitions in the effective
regime are now
\begin{align}
c= & \frac{\gamma a^{\prime}\mathring{v}^{\frac{1}{3}}\text{sgn}\left(p\right)}{f_{c}}, & \left|p\right|= & a^{2}\mathring{v}^{\frac{2}{3}}.\label{eq:c-p-def-modified}
\end{align}
Equations~\eqref{eq:c-prime-eff-general} and 
\eqref{eq:p-prime-eff-general} once again lead to the Friedmann equations; 
however, in this case they acquire modifications due to GUP effects. 
Following the same procedure as in the classical case, but using the 
definitions in \eqref{eq:c-p-def-modified}, we obtain
\begin{equation}
    H^{2}=  \frac{\kappa}{3}\rho f^{2}_{c}= \frac{\kappa}{3}\rho\left(1+\frac{\kappa\gamma^{2}}{3}\beta_{c}\rho|p|\right)^{2},\label{eq:Fried-1-eff}
\end{equation}
and
\begin{align}
\frac{a^{\prime\prime}}{a}= -\frac{\kappa}{6}\left(\rho+3P\right)&f^{2}_{c}\pm\text{sgn}\left(p\right)\sqrt{\frac{\kappa}{3}}\rho^{\frac{1}{2}}f^{\prime}_{c}\nonumber \\
= -\frac{\kappa}{6}\left(\rho+3P\right)&-\frac{8\kappa^{2}}{9}\beta_{c}\gamma^{2}\rho^{2}\left|p\right|-\frac{2\kappa^{3}}{9}\beta^{2}_{c}\gamma^{4}\rho^{3}\left|p\right|^{2}\nonumber\\
&+\text{sgn}\left(p\right)\frac{2\kappa^{2}}{9}\left(\gamma^{2}\rho^{2}\left|p\right|^{2}+\frac{\kappa}{3}\beta_{c}\gamma^{4}\rho^{3}\left|p\right|^{3}\right)\frac{d\beta_{c}}{dp}\label{eq:Fried-2-eff}
\end{align}
where we have used $f_{c}=1+\beta_{c}c^{2}$ regardless of whether
$\beta_{c}$ depends on $p$ or not, and 
\begin{align}
f^{\prime}_{c}=\frac{d}{d\tau}\left(1+\beta_{c}c^{2}\right)= & \frac{\partial f_{c}}{\partial c}c^{\prime}+\frac{\partial f_{c}}{\partial\beta_{c}}\frac{d\beta_{c}}{dp}p^{\prime}\nonumber \\
= & \pm\frac{2\kappa\gamma^{2}}{3}\sqrt{\frac{\kappa}{3}}\rho^{\frac{3}{2}}|p|f_{c}\left(-2\text{sgn}\left(p\right)\beta_{c}+|p|\frac{d\beta_{c}}{dp}\right).
\end{align}
The three equations (\ref{eq:EoM-diff-eff}), (\ref{eq:Fried-1-eff})
and (\ref{eq:Fried-2-eff}) serve as three master equations that we
will study for subcases in what follows. 

\subsubsection{Subcase 1: Constant $\beta_{c}$ \label{subsec:beta-constant}}

By considering a constant deformation parameter $\beta_{c}$, Eq.
(\ref{eq:EoM-diff-eff}) becomes 
\begin{equation}
\frac{dp}{d\phi}=\pm\sqrt{\frac{2\kappa}{3}}\left(\left|p\right|+\text{sgn}\left(\beta_{c}\right)\ell^{2}_{c}\frac{p^{2}_{\phi}}{\left|p\right|}\right),\qquad\ell_{c}=\sqrt{\frac{\left|\beta_{c}\right|\kappa}{6}}\gamma,\label{eq:EoM-beta-constant}
\end{equation}
where $\ell_{c}$ is a length scale associated to this model. We have
separated $\beta_{c}$ from its sign since different signs of $\beta_{c}$
carry important distinct physical consequences \cite{Fragomeno:2024tlh, Gingrich:2024mgk}.The
solution to this equation is
\begin{equation}
p\left(\phi\right)=\pm\sqrt{C_{1}e^{\pm\text{sgn}\left(p_{0}\right)\sqrt{\frac{8\kappa}{3}}\phi}-\text{sgn}\left(\beta_{c}\right)\ell^{2}_{c}p^{2}_{\phi}}\label{eq:p-phi-beta-const}
\end{equation}
where $C_{1}=p\left(0\right)^{2}+\text{sgn}\left(\beta_{c}\right)\ell^{2}_{c}p^{2}_{\phi}$
is a constant of integration. The above can be rewritten in a more
insightful way as
\begin{equation}
V=\left(C_{1}e^{\pm\text{sgn}\left(p_{0}\right)\sqrt{\frac{8\kappa}{3}}\phi}-\text{sgn}\left(\beta_{c}\right)\mathscr{V}^{\frac{4}{3}}_{c}\right)^{\frac{3}{4}},\label{eq:V-phi-beta-const}
\end{equation}
where
\begin{equation}
\mathscr{V}_{c}=\left(\ell_{c}p_{\phi}\right)^{\frac{3}{2}}=\left(\sqrt{\frac{\left|\beta_{c}\right|\kappa}{6}}\gamma p_{\phi}\right)^{\frac{3}{2}}.
\end{equation}
This defines a characteristic volume scale associated with $\ell_{c}$ 
and the constant of motion $p_{\phi}$. Here we have used 
Eqs.~\eqref{eq:rho-p-class} and \eqref{eq:p-vs-V}. This corresponds to 
the minimum volume that a Universe with $\beta_{c}<0$ asymptotically 
attains during the contraction phase, or from which it emerges in the 
expansion phase, instead of reaching $V=0$ as in the classical or 
$\beta_{c}>0$ case. 

This minimum volume depends on both the energy density and the quantum 
deformation parameter. As before, the $+$ and $-$ signs in 
\eqref{eq:V-phi-beta-const} correspond to expanding and contracting 
branches, respectively. 

\begin{figure}[htb!] 
  \centering
  \includegraphics[width=1.0\linewidth]{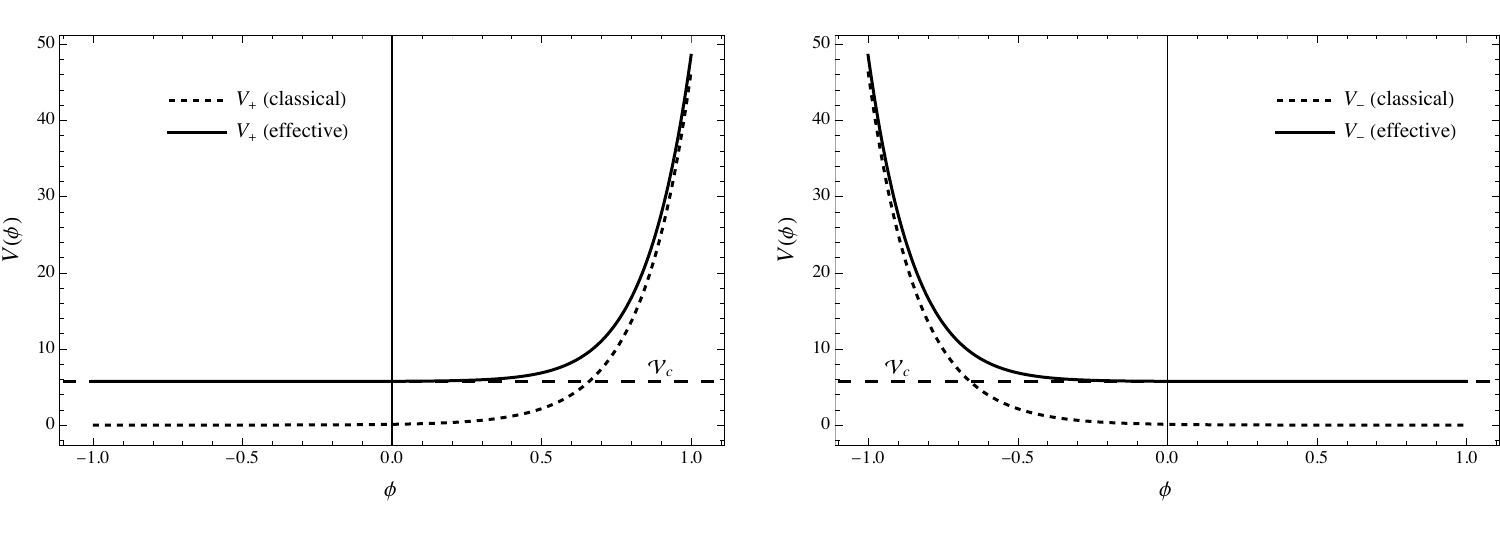} 
  \caption{Behavior of the solutions for \(V_{\pm}\) as functions of \(\phi\) in the classical vs. the effective regimes. 
The solid curve in the left panel represents the expansion of the 
Universe, while the one in the right panel corresponds to its 
contraction. The dashed curve denotes the classical cosmological 
dynamics.}
  \label{fig:p_eff_vs_phi_compl} 
\end{figure}

This behavior is illustrated in Fig.~\ref{fig:p_eff_vs_phi_compl}, where 
the effective solution (solid curve) is compared with the classical one 
(dashed curves). The left panel shows the expansion phase, while the 
right panel corresponds to contraction. The values of the constants and 
parameters have been chosen to clearly highlight the differences between 
the classical and effective cases.

To study the modified behavior of the universe in this case, we note
that
\begin{equation}
\lim_{\phi\to-\infty}V=\mathscr{V}_{c}\left(-\text{sgn}\left(\beta_{c}\right)\right)^{\frac{3}{4}}.\label{eq:lim-beta-const-phi--inf}
\end{equation}
If $\beta_{c}>0$, the above limit does not exist. This is because,
e.g., for a contracting universe, the term inside the parenthesis
in (\ref{eq:V-phi-beta-const}) can actually becomes zero and the
universe will have $V=0$ for 
\begin{equation}
C_{1}e^{\sqrt{\frac{8\kappa}{3}}\phi}=\mathscr{V}^{\frac{4}{3}}_{c}
\end{equation}
Thus, for $\beta_{c}>0$, the singularity is not resolved. Instead,
a state of zero volume is reached at a finite clock time in the contracting
phase (or at some time in the past for the expanding case), given
by:
\begin{equation}
\phi=\sqrt{\frac{3}{8\kappa}}\ln\left(\frac{\mathscr{V}^{\frac{4}{3}}_{c}}{C_{1}}\right).
\end{equation}
For $\beta_{c}<0$, however, the limit \eqref{eq:lim-beta-const-phi--inf} 
shows that $V$ admits a horizontal asymptote at $\mathscr{V}_{c}$. 
Consequently, the Universe possesses a finite minimum volume 
$\mathscr{V}_{c}$ at early times (or at late times in the contracting 
branch), as illustrated in Fig.~\ref{fig:p_eff_vs_phi_compl}. In this
picture, the universe ``coasts'' out of a non-singular quantum state
of constant volume of size $\mathscr{V}_{c}$ in the infinite past
($\phi\rightarrow-\infty$) \cite{Ellis:2003qz, Ellis:2003ue}. This
is in contrast to the well-known classical solution (\ref{eq:V-phi-Class})
in which the volume vanishes for $\phi\to-\infty$, as well as approaches
like loop quantum cosmology (LQC) where there is a bounce.

On the other hand, at large times we obtain
\begin{equation}
V\left(\phi\to\infty\right)=C^{\frac{3}{4}}_{1}e^{\sqrt{\frac{3\kappa}{2}}\phi}-\text{sgn}\left(\beta_{c}\right)\frac{3}{4}C^{-\frac{1}{4}}_{1}\mathscr{V}^{4/3}_{c}e^{-\sqrt{\frac{\kappa}{6}}\phi}+\mathcal{O}\left(e^{-5\sqrt{\frac{\kappa}{6}}\phi}\right).
\end{equation}
The first leading term is precisely the classical behavior (\ref{eq:V-phi-Class})
for $C^{\frac{3}{4}}_{1}=V_{0}$. The second term is the leading quantum
gravity term. The leading correction decays as $e^{-\sqrt{\kappa/6}\phi}$
and is suppressed relative to the classical term by a factor $e^{-\sqrt{8\kappa/3}\phi}$,
i.e.,

\[
\frac{\delta V}{V_{\text{class}}}=-\frac{3\text{sgn}\left(\beta_{c}\right)}{4}\frac{\mathscr{V}^{4/3}_{c}}{C_{1}}e^{-\sqrt{\frac{8\kappa}{3}}\phi}\rightarrow0\quad(\phi\rightarrow\infty)
\]
which vanishes rapidly. 

Furthermore, from (\ref{eq:EoM-beta-constant}), one can see that
for $\beta_{c}<0$ and expanding branch $\text{sgn}\left(p\right)>0$
and choosing the $+$ sign in $\pm$, the fixed point $p_{*}$ is
\begin{equation}
\frac{dp}{d\phi}=0\Rightarrow\left|p_{*}\right|=\ell_{c}p_{\phi}\Rightarrow V_{*}=\mathscr{V}_{c}.
\end{equation}
Perturbing the universe around this point 
\begin{equation}
p=p_{*}+\delta p,\quad|\delta p|\ll p_{*}
\end{equation}
yields to first order in $\delta p$
\begin{equation}
\frac{d\left(\delta p\right)}{d\phi}=\sqrt{\frac{8\kappa}{3}}\delta p\Rightarrow\delta p\propto e^{\sqrt{\frac{8\kappa}{3}}\phi}.
\end{equation}
Hence, for the expanding branch, perturbations grow exponentially
as $\phi$ increases and the universe is repelled from the fixed point,
showing it is unstable, since the Lyapunov exponent is 
\begin{equation}
\lambda_{1}=\sqrt{\frac{8\kappa}{3}}.
\end{equation}
In the contracting case, perturbations decay as $\phi\rightarrow-\infty$,
meaning the universe only approaches $\mathscr{V}_{c}$ asymptotically,
never departing from it in finite clock time, consistent with the
solution structure.

The Friedmann equations (\ref{eq:Fried-1-eff}) and (\ref{eq:Fried-2-eff})
in this case become
\begin{align}
\left(\frac{a^{\prime}}{a}\right)^{2}= & \frac{\kappa}{3}\rho\left(1+\beta_{c}\frac{\rho}{\rho^{(0)}_{c}}\right)^{2},\label{eq:Fried-1-beta-constant}\\
\frac{a^{\prime\prime}}{a}= & -\frac{\kappa}{6}\left(\rho+3P\right)-\frac{8\kappa^{2}}{9}\beta_{c}\gamma^{2}\rho^{2}\left|p\right|-\frac{2\kappa^{3}}{9}\beta^{2}_{c}\gamma^{4}\rho^{3}\left|p\right|^{2},\label{eq:Fried-2-beta-constant}
\end{align}
where
\begin{equation}
\rho^{(0)}_{c}=\left(\frac{\kappa\gamma^{2}}{3}\left|p\right|\right)^{-1}=\left(\frac{\kappa\gamma^{2}}{3}V^{\frac{2}{3}}\right)^{-1}=\left(\frac{\kappa\gamma^{2}}{3}a^{2}\mathring{v}^{\frac{2}{3}}\right)^{-1}.\label{eq:crit-rho-beta-const}
\end{equation}
We see from the (\ref{eq:Fried-1-beta-constant}) that if $\beta_{c}<0$,
there is a critical density (\ref{eq:crit-rho-beta-const}) for which
the contraction stops. This happens when the energy density is $\rho=-\frac{\rho^{(0)}_{c}}{\beta_{c}}$.
Since the right hand side of (\ref{eq:Fried-1-beta-constant}) is
always strictly nonnegative, no bounce will happen in line with what
we mentioned when discussing the solution (\ref{eq:V-phi-beta-const}).
However, the value of $\rho^{(0)}_{c}$ depends on the size of the
universe and is not a universal one in the $\beta_{c}$ constant case.
Also there is a fiducial anomaly: from the last term in (\ref{eq:crit-rho-beta-const})
it is clear that the physical density at which quantum gravity effects
kick in depends on the arbitrary choice of the fiducial cell volume
$\mathring{v}$. This is a severe unphysical artifact and is analogous
to the $\mu_{0}$ scheme in LQC. This anomaly is present also in the
second Friedmann equation (\ref{eq:Fried-2-beta-constant}), where
the last two terms depend on $\left|p\right|$ and hence on the fiducial
volume $\mathring{v}$. As we will see, this is remedied by introducing the improved scheme, which we consider in the next section.

\subsubsection{Subcase 2: $\beta_{c}\propto\frac{1}{p}$ and beyond\label{subsec:improved-beta}}

In this case, we consider a momentum dependent function 
\begin{equation}
\bar{\beta}_{c}(p)=\beta_{c}\frac{\ell^{2}_{p}}{\left|p\right|}
\end{equation}
with $\ell_{p}$ being the Planck length, such that the deformation
function $f_{c}$ takes the form 
\begin{equation}
f_{c}=1+\bar{\beta}_{c}\left(p\right)c^{2}=1+\beta_{c}\ell^{2}_{p}\frac{c^{2}}{\left|p\right|}.\label{eq:f-c-improved}
\end{equation}
The master EoM Eq. (\ref{eq:EoM-diff-eff}) becomes
\begin{align}
\frac{dp}{d\phi}= & \pm\sqrt{\frac{2\kappa}{3}}\left|p\right|\left(1+\bar{\beta}_{c}\left(p\right)c^{2}\right)\nonumber \\
= & \pm\sqrt{\frac{2\kappa}{3}}\left(\text{sgn}\left(p\right)p+\frac{\kappa}{6}\ell^{2}_{p}\beta_{c}\gamma^{2}\frac{p^{2}_{\phi}}{p^{2}}\right),\label{eq:EoM-beta-p}
\end{align}
where we have used (\ref{eq:c-from-constraint}). The solution to
this equation reads
\begin{align}
p\left(\phi\right) & =\text{sgn}\left(p_{0}\right)\left(C_{2}e^{\pm\text{sgn}\left(p_{0}\right)\sqrt{6\kappa}\phi}-\kappa\ell^{2}_{p}\frac{\beta_{c}\gamma^{2}}{6}p^{2}_{\phi}\right)^{\frac{1}{3}},\label{eq:sol-p-eff-fc-improved}\\
V\left(\phi\right) & =\left(C_{2}e^{\pm\text{sgn}\left(p_{0}\right)\sqrt{6\kappa}\phi}-\text{sgn}\left(\beta_{c}\right)\ell^{2}_{p}\mathscr{V}^{\frac{4}{3}}_{c}\right)^{\frac{1}{2}}.\label{eq:sol-V-eff-fc-improved}
\end{align}
One can see that again there is a horizontal asymptote for 
\begin{equation}
\lim_{\phi\to-\infty}V=\sqrt{-\text{sgn}\left(\beta_{c}\right)}\ell_{p}\mathscr{V}^{\frac{2}{3}}_{c}
\end{equation}
which is only real if $\beta_{c}<0$. This universe has also the qualitative
behavior as the one with constant $\beta_{c}$ but with an important
difference: the elimination of the fiducial anomaly discussed in the
previous section as we will see below. Moreover, in this case, again
there are two branches of the universe: the expanding branch emerges
from an asymptotically constant-volume phase with a volume 
\begin{equation}
V_{\text{min}}=\ell_{p}\mathscr{V}^{\frac{2}{3}}_{c}\label{eq:V-min-bar}
\end{equation}
and the contracting universe that smoothly asymptotes to a final constant
volume state with the same volume. There is no bounce in this subcase
either.

The dynamics described by \eqref{eq:sol-V-eff-fc-improved} are similar 
to those of \eqref{eq:V-phi-beta-const}, with the key difference that 
the minimum volume is now given by \eqref{eq:V-min-bar}. Consequently, 
the behavior of \eqref{eq:sol-V-eff-fc-improved} resembles that shown 
in Fig.~\ref{fig:p_eff_vs_phi_compl}.

Moreover, in this case, we obtain
\begin{equation}
V\left(\phi\to\infty\right)=\sqrt{C_{2}}e^{\sqrt{\frac{3}{2}\kappa}\phi}-\text{sgn}\left(\beta_{c}\right)\frac{1}{2\sqrt{C_{2}}}\ell^{2}_{p}\mathscr{V}^{4/3}_{c}e^{-\sqrt{\frac{3}{2}\kappa}\phi}+\mathcal{O}\left(e^{-3\sqrt{\frac{3}{2}\kappa}\phi}\right).
\end{equation}
Again the first leading term is the classical behavior (\ref{eq:V-phi-Class})
for $\sqrt{C_{2}}=V_{0}$. The leading quantum gravity term, decays
as $e^{-\sqrt{\frac{3}{2}\kappa}\phi}$ and is suppressed relative
to the classical term by

\[
\frac{\delta V}{V_{\text{class}}}=-\frac{\text{sgn}\left(\beta_{c}\right)\ell^{2}_{p}\mathscr{V}^{4/3}_{c}}{2C_{2}}e^{-\sqrt{6\kappa}\phi}\rightarrow0\quad(\phi\rightarrow\infty),
\]
which again vanishes rapidly. On the other hand the corrections in
this case decays as $e^{-\sqrt{6\kappa}\phi}$ versus $e^{-\sqrt{\frac{8}{3}\kappa}\phi}$
in $\beta_{c}=$constant case. This means that the improved scheme
returns to classical behavior significantly faster in clock time.

From (\ref{eq:EoM-beta-p}), one can see that for $\beta_{c}<0$ and
expanding branch $\text{sgn}\left(p\right)>0$ and choosing the $+$
sign in $\pm$, the fixed point $p_{*}$ is
\begin{equation}
\frac{dp}{d\phi}=0\Rightarrow p^{3}_{*}=\ell^{2}_{p}\ell^{2}_{c}p^{2}_{\phi}.
\end{equation}
Perturbing the universe around this point 
\begin{equation}
p=p_{*}+\delta p,\quad|\delta p|\ll p_{*}
\end{equation}
yields to first order in $\delta p$
\begin{equation}
\frac{d\left(\delta p\right)}{d\phi}=\sqrt{6\kappa}\delta p\Rightarrow\delta p\propto e^{\sqrt{6\kappa}\phi}.
\end{equation}
This is qualitatively the same unstable behavior as the $\beta_{c}$
case, showing that the initial phase of the universe is unstable and
is repelled ``outwards''. But the Lyapunov exponent is now 
\begin{equation}
\lambda_{2}=\sqrt{6\kappa}
\end{equation}
which shows that the universe is repelled out of the constant-volume
phase faster than the constant $\beta_{c}$ case.

The Friedmann equations in this case can be found from (\ref{eq:Fried-1-eff})
and (\ref{eq:Fried-2-eff}) by use of $\bar{\beta}_{c}$, as 
\begin{align}
H^{2}= & \frac{\kappa}{3}\rho\left(1+\beta_{c}\frac{\rho}{\bar{\rho}_{c}}\right)^{2},\label{eq:Fried-1-beta-p}\\
\frac{a^{\prime\prime}}{a}= & -\frac{\kappa}{6}\left(\rho+3P\right)-\frac{10\kappa^{2}\ell^{2}_{p}}{9}\beta_{c}\gamma^{2}\rho^{2}-\frac{8\kappa^{3}\ell^{4}_{p}}{27}\beta^{2}_{c}\gamma^{4}\rho^{3},\label{eq:Fried-2-beta-p}
\end{align}
where the maximum, and now invariant, energy density is
\begin{equation}
\bar{\rho}_{c}=\frac{3}{\kappa\ell^{2}_{p}\gamma^{2}}.\label{eq:rho-c-bar}
\end{equation}
Once again the first Friedmann equation shows that for $\beta_{c}<0$,
the term inside the parenthesis in the RHS of (\ref{eq:Fried-1-beta-p})
becomes zero for $\rho=-\frac{\bar{\rho}_{c}}{\beta_{c}}$. However,
in this case, neither $\bar{\rho}_{c}$ nor the RHS of Friedmann equations
depend on $p$ or equivalently on the fiducial volume. The fiducial
anomaly is resolved. As expected, although the qualitative behavior
is the same as the constant $\beta_{c}$ case, the improved scheme
has resolved the unphysical $\mathring{v}$-dependent behavior. 

We can see that both these scenarios point to a no-bounce scenario
(unlike LQC) without any singularity at the beginning or end of the
universe, where the asymptotic phase of the universe (asymptotic past
in expanding case and asymptotic future in contracting case) is a
constant-volume state. In particular, the expanding case is similar
to emergent universe scenarios where instabilities of perturbations
make the universe transition from a constant-volume phase to an expanding
one.

One could in principle consider a general power-law deformation $\bar{\beta}(p)\propto|p|^{n}$.
However, observing the generic structure of Eq. (\ref{eq:Fried-1-eff}),
the quantum correction term scales as $\rho|p|^{n+1}$. To strictly
eliminate the fiducial anomaly and ensure quantum gravity effects
manifest at a universal, invariant energy density, we are uniquely
constrained to choose $n=-1$ for FLRW. 

\subsection{Case 2: Deforming Both the Geometry and Matter Sectors\label{subsec:Deform-both}}

We will now consider the case that in addition to improved GUP modification
of the geometry in subcase \ref{subsec:improved-beta}, Eq. (\ref{eq:PB-orig-cp-GUP}) with
(\ref{eq:f-c-improved}), the algebra of the matter sector is also
modified as
\begin{equation}
\left\{ \phi,p_{\phi}\right\} =f_{m}\label{eq:matter-PB-eff}
\end{equation}
with
\begin{equation}
f_{m}\left(\beta_{m},p_{\phi}\right)=\left(1+\beta_{m}p^{2}_{\phi}\right)\label{eq:fm-p-phi}
\end{equation}
The effective EoM now become
\begin{align}
c^{\prime}= & -\frac{2}{\gamma}\frac{\text{sgn}\left(p\right)}{\left|p\right|^{\frac{1}{2}}}c^{2}f_{c},\\
p^{\prime}= & \frac{2}{\gamma}\left|p\right|^{\frac{1}{2}}cf_{c},\\
\phi^{\prime}= & \frac{p_{\phi}}{\left|p\right|^{\frac{3}{2}}}f_{m},\\
p^{\prime}_{\phi}= & 0,
\end{align}
which leads to 
\begin{equation}
\frac{dp}{d\phi}=\pm\sqrt{\frac{2\kappa}{3}}\left|p\right|\frac{f_{c}}{f_{m}}.\label{eq:EoM-fc-fm}
\end{equation}
Since $p_{\phi}$ is still a constant of motion, $f_{m}$ is also
a constant and thus we can write
\begin{equation}
\frac{dp}{d\left(\frac{\phi}{f_{m}}\right)}=\pm\sqrt{\frac{2\kappa}{3}}\left|p\right|f_{c}.
\end{equation}
This shows that deforming matter on top of deforming geometry is equivalent
to rescaling the clock variable $\phi\to\frac{\phi}{f_{m}}$ in the
case of just deforming the geometry.

Moreover, since $f_{m}$ is a constant, and the new deformation of
the matter algebra (\ref{eq:EoM-fc-fm}) does not affect the EoM of
$c,\,p$ since $c^{\prime},\,p^{\prime}$ does not depend on $\phi$,
the new EoM (\ref{eq:EoM-fc-fm}) can be written as
\begin{equation}
\frac{dp}{d\phi}=\frac{1}{f_{m}}\left(\frac{dp}{d\phi}\right)_{f_{m}=1}
\end{equation}
where $\left(\frac{dp}{d\phi}\right)_{f_{m}=1}$ refer to the EoM
of the model in which just the geometry algebra is deformed with $\beta_{c}(p)=\frac{\beta_{c}\ell^{2}_{p}}{p}$,
Eq. (\ref{eq:EoM-beta-p}). The solution to this equation is
\begin{align}
p\left(\phi\right) & =\text{sgn}\left(p_{0}\right)\left(C_{2}e^{\pm\frac{\text{sgn}\left(p_{0}\right)}{f_{m}}\sqrt{6\kappa}\phi}-\kappa\ell^{2}_{p}\frac{\beta_{c}\gamma^{2}}{6}p^{2}_{\phi}\right)^{\frac{1}{3}},\label{eq:sol-p-eff-fm-improved}\\
V\left(\phi\right) & =\left(C_{2}e^{\pm\frac{\text{sgn}\left(p_{0}\right)}{f_{m}}\sqrt{6\kappa}\phi}-\text{sgn}\left(\beta_{c}\right)\ell^{2}_{p}\mathscr{V}^{\frac{4}{3}}_{c}\right)^{\frac{1}{2}}.\label{eq:sol-V-eff-fm-improved}
\end{align}
As expected the solutions are the same as (\ref{eq:sol-p-eff-fc-improved})
and (\ref{eq:sol-V-eff-fc-improved}) with $\phi\to\frac{\phi}{f_{m}}$.

Because $c^{\prime}$ and $p^{\prime}$ do not depend on $\phi$,
the Friedmann equations remain the same as (\ref{eq:Fried-1-beta-p})
and (\ref{eq:Fried-2-beta-p}). Furthermore the minimum volume $\ell_{p}\mathscr{V}^{\frac{2}{3}}_{c}$
and the maximum (critical) invariant energy density $\frac{3}{\kappa\ell^{2}_{p}\gamma^{2}}$
remain unchanged and are the same as before in (\ref{eq:V-min-bar})
and (\ref{eq:rho-c-bar}), respectively. Thus, the matter deformation
cannot force a bounce. The universe remains an emergent one, coasting
out of the exact same geometric minimum constant-volume state as before.

The rate of evolution is now governed by a modified Lyapunov exponent
\[
\lambda_{3}=\frac{\sqrt{6\kappa}}{f_{m}}=\frac{\lambda_{2}}{f_{m}}
\]
If one considers $\beta_{\phi}>0$, which we do here, then $f_{m}>1$
and thus $\lambda_{3}<\lambda_{2}$. Thus, the quantum matter deformation
acts as a relational ``time dilation'' (relative to the scalar field
clock $\phi$ without deformation). The universe is repelled away
from the emergent constant-volume state more slowly than it would
be if only geometry were deformed. The higher the initial matter $p_{\phi}$,
the stronger this relational slowing effect becomes.

Let us briefly comment on the case where the matter deformation is
in the form 
\begin{equation}
\left\{ \phi,p_{\phi}\right\} =f_{m}(\phi)\coloneqq1+\beta_{\phi}\phi^{2}.\label{eq:fm-phi}
\end{equation}
The solutions to the geometry sector EoM lead to
\begin{equation}
V(\phi)=\left(C_{2}\exp\left[\pm\operatorname{sgn}\left(p_{0}\right)\sqrt{\frac{6\kappa}{\beta_{\phi}}}\tan^{-1}\left(\sqrt{\beta_{\phi}}\phi\right)\right]-\operatorname{sgn}\left(\beta_{c}\right)\ell^{2}_{p}\mathscr{V}^{\frac{4}{3}}_{c}\right)^{\frac{1}{2}}
\end{equation}
which means
\begin{equation}
\lim_{\phi\rightarrow\infty}V(\phi)=\left(C_{2}\exp\left[\pm\operatorname{sgn}\left(p_{0}\right)\sqrt{\frac{6\kappa}{\beta_{\phi}}}\frac{\pi}{2}\right]-\operatorname{sgn}\left(\beta_{c}\right)\ell^{2}_{p}\mathscr{V}^{\frac{4}{3}}_{c}\right)^{\frac{1}{2}}\eqqcolon V_{\max},
\end{equation}
and hence the universe stops expanding after it reaches $V_{\text{max}}$.
However, the issue in this case is that the $\phi$ EoM becomes
\begin{equation}
\phi^{\prime}=\frac{p_{\phi}}{V}\left(1+\beta_{\phi}\phi^{2}\right),
\end{equation}
and this shows that as $\phi\to\infty$ and $V\to V_{\text{max}}$,
the clock rate $\phi^{\prime}$ goes to infinity, clock's speed to
run away to infinity and this happens at a finite proper time equal
to
\begin{equation}
\Delta\tau\approx\int^{\infty}_{\phi_{0}}\frac{V_{\max}}{p_{\phi}}\frac{d\phi}{\beta_{\phi}\phi^{2}}=\frac{V_{\max}}{p_{\phi}\beta_{\phi}\phi_{0}}.
\end{equation}
The universe does not gently coast forever. Instead the physical clock's
rate runs away to infinity and ceases to be a valid mathematical parameter
or a physical clock. The deformation (\ref{eq:fm-phi}) also amount
to a nonlinear transformation of $\phi$ (compared to the case that
matter algebra is not modified at all), and this ruins the shift symmetry
$\phi\to\phi+C$ of the scalar field. For these reasons, we conclude
that the deformation (\ref{eq:fm-p-phi}) is the suitable modification
to the matter sector. 

\section{Conclusion\label{Sec:Conclusion}}

We have studied a GUP-deformed effective cosmology for a spatially
flat FLRW universe coupled to a massless scalar field, working within
the Ashtekar-Barbero phase space formulation. By deforming the classical
Poisson algebra while leaving the Hamiltonian constraint unchanged,
we derived effective equations of motion and modified Friedmann equations
across three progressively richer scenarios.

In the geometry-only deformation with a constant GUP parameter $\beta_{c}$,
we found that the sign of the deformation parameter is physically
decisive. For $\beta_{c}>0$ the singularity persists and is in fact
reached at a finite value of the scalar clock, earlier than in the
classical case. For $\beta_{c}<0$ the singularity is resolved: the
universe asymptotes to a non-singular state of constant volume $\mathscr{V}_{c}$
in the infinite past, with no bounce. A fixed-point analysis confirms
this state is an unstable fixed point with Lyapunov exponent $\lambda_{1}=\sqrt{8\kappa/3}$,
so that small perturbations grow and drive the universe into classical
expansion --- precisely the mechanism of the emergent universe scenario.
A serious shortcoming of this case is a fiducial-cell anomaly in the
critical density, analogous to the $\mu_{0}$ scheme in LQC, rendering
the scale at which quantum effects appear dependent on the arbitrary
fiducial volume.

To rectify this, we introduced an improved scheme by promoting the
geometric deformation parameter $\beta_{c}$ to a momentum-dependent
function, $\bar{\beta}_{c}(p)\propto\ell^{2}_{p}/|p|$. This improved
scheme eliminates this anomaly entirely. We showed that this is the
unique power-law choice $\bar{\beta}_{c}\propto|p|^{n}$ for which
the quantum correction term in the first Friedmann equation is independent
of the fiducial cell, singling out $n=-1$ as the physically preferred
scheme. The qualitative picture --- emergent universe, no bounce,
asymptotic constant-volume phase --- is preserved, but with a universal,
invariant critical density $\bar{\rho}_{c}=3/(\kappa\ell^{2}_{p}\gamma^{2})$
and a larger Lyapunov exponent $\lambda_{2}=\sqrt{6\kappa}>\lambda_{1}$.
This means the improved scheme not only resolves the fiducial anomaly
but also predicts a faster transition from the emergent phase to classical
expansion, a concrete quantitative advantage over the constant-$\beta_{c}$
case. The asymptotic expansion at large clock times confirms exponentially
fast recovery of the classical trajectory, with the leading correction
decaying as $e^{-\sqrt{6\kappa}\,\phi}$ versus $e^{-\sqrt{8\kappa/3}\,\phi}$
in the constant case.

When the matter sector algebra is also deformed by a constant positive
parameter $\beta_{m}$, we established a clean structural result:
since $p_{\phi}$ is conserved, $f_{m}$ is a constant, and matter
deformation is exactly equivalent to a rescaling of the scalar clock
$\phi\to\phi/f_{m}$. As a consequence, the Friedmann equations, the
minimum volume, and the critical density are entirely unaffected by
the matter deformation. The only physical effect is a relational time
dilation: the Lyapunov exponent becomes $\lambda_{3}=\lambda_{2}/f_{m}<\lambda_{2}$,
meaning the universe exits the emergent phase more slowly relative
to the scalar clock when $\beta_{m}>0$. These results provide a rigorous
framework for emergent universe scenarios and highlight the fundamental
necessity of momentum-dependent deformation schemes in effective quantum
cosmology. 


\acknowledgments{
S.R. acknowledges the support of the Natural Sciences and Engineering
Research Council of Canada (NSERC). W. Y thanks the CONAHCyT Grant CBF2023-2024-2923 ``Implications of the Generalized Uncertainty Principle (GUP) in Quantum Cosmology, Gravitation, and its Connection with Non-extensive Entropies'', and is supported by SECIHTI/Estancias Posdoctorales por M\'exico.
}












\bibliographystyle{jhep}





\bibliography{mainbib}

\end{document}